# E-learning and use of computer in forensic field

**Alina Oana Zamosteanu**
"Tibiscus" University, Timisoara, Romania

**ABSTRACT.** In the Romanian penitentiary establishments and those of other European countries one talks about the formal and informal education. Since the correspondence education system, important steps have been made towards the e-learning didactics which are reflected by the modern teaching methods (through ICT) used even by the penitentiary system. The Moodle platform, the web site with certain specific, and the forum represent the means used as an interface between the educator and the student, their benefits being clearly demonstrated.

**Introduction**

In order to conduct some effective educational and therapeutic activities in the rehabilitation centres, in the youth penitentiaries and in the adult prisons, the applied model is the medical one: investigation-diagnosis-treatment-monitoring-evaluation.

During the execution of the sentence, each person benefits of an educational program for the period of time to be spent in the institution. The educational program takes into account not the committed deed, but the particular psycho- individual features of each person. In order to accomplish this, the first step is the observation stage – gathering relevant information about his/her history, identifying the psycho-behavioural profile, and the interests and skills of each person. A file is made up including the personalized intervention program. Based on this evaluation, every person deprived of freedom is involved in various programs and activities from one of the following areas:

1. The educational area and the area of social skills development which aim at the formation of daily life habits, at the knowledge development and at the capacity of using this information in everyday life.

2. The area of the psycho-behavioural balancing and optimization intends to compensate for the imbalances in the personality. It also aims to adequate the self image and to create the personal resources awareness, therefore contributing to the improvement of interpersonal relationships.

3. The instructive-formative area (education and professional qualification) – the initiation / completion of education in accordance with the public education system, as well as the professional training and qualification according to the individual interest, to the individual skills and, if possible, to the labour market trends.

4. The area of artistic and sports-leisure activities – for the maintenance of an age-appropriate physical condition and for developing the team spirit and fair play, for discovering and valuing the artistic inclinations and skills in leisure activities.

## 1. Programs and interventions in penitentiary using ICT and e-learning methods

The National Administration of Penitentiaries has planned an educational policy and psycho-social assistance program, primarily oriented towards the following areas:
- school instruction - regardless of the age and level of training before entry into prison. It is conceived as education for adults, yet in the form of primary and secondary education as a second chance. Under certain conditions one may benefit of university instruction.
- professional training - organized as schools of arts and crafts or as conversion or retraining courses conducted in collaboration with the Ministry of Education and Research and with the County Employment Agencies.
- the development of skills and social competences for the acquirement of a pro-social conduct and of principles for a normal life, in line with the social norms in order to better manage their life in some areas (health, family) or to overcome negative aspects of life (alcoholism, drug addiction);
- physical education and sports are useful means of leisure and they are much appreciated by the persons deprived of liberty due to the human desire of being daily active;

- the library – it is an important source of unstructured education that contributes to the personal development. At the same time, it is a source of entertainment, but also a centre of cultural development. The persons deprived of liberty have weekly access to the library that exists in each centre of detention;
- educational and cultural activities - they involve both the active participation of the persons deprived of liberty (artistic expression - painting, modelling, sculpture), and their passive participation (watching movies, plays, attending lectures, etc.);
- psychological assistance – it consists of an initial psychological evaluation, performed by a clinical interview and a kit of psychological tests. It is a key component of the educational and therapeutic intervention plan. Psychological intervention in case of emergency is also provided for the people deprived of liberty when accusing an acute existential situation, expressed in behavioural manifestations of pathogenic origin (increased anxiety, psychomotor agitation, depression, etc.). The psychological intervention aims to improve the psychic functioning (the control of impulsiveness and frustration tolerance, the ability of objective self-assessment), but also to improve the perception of reality (tolerance of ambivalence, more human and consistent relationships);
- the social assistance – it requires the knowledge and understanding of the social history of the person deprived of liberty (family of origin, family dynamics, social and economic structure). This data is required to design the plan of educational and psychotherapeutic intervention. Also the specialists working in penitentiaries offer family counselling to improve the relations between family members and to involve the family in the process of social reintegration.

In parallel with these activities a coherent methodology has been created to work with the juvenile, young or adult people in detention. Also the necessary documents for an effective rehabilitation intervention have been completed:

1. The guide for good practice for the staff working with the juvenile, young and adult people in detention, which contains principles, procedures, organizational arrangements for a detention in a dynamic, constructive and, where possible, individual way. It addresses all the staff categories and it intends to set the fundaments for a multidisciplinary team able to provide a comprehensive rehabilitation action.

2. The educational project of the re-educational centres, the guide of educational / therapeutic activities to be carried out with the juvenile prisoners; guides of programs for adults (persons deprived of liberty)
3. The guide for the youth who go to prison - which aims to facilitate the adaptation of new imprisoned juveniles to the penitentiary rules, to diminish the shock of imprisonment, and to present the activities offer.

The educational and psychosocial assistance specific to the Romanian prison system presents many similarities with the models promoted in other European states. For example, Spain has a long tradition with regard to education in prison. Historically, it is one of the first countries that adopted a human perspective in terms of criminal doctrine. The compulsiveness of the existence of schools and teachers in penitentiaries was stipulated by an ordinance in 1834. Education is considered the main route to social reintegration, which is why emphasis is on a modern and integrated perspective of specific adult learning. Currently, the penitentiaries in Spain are developing over 2000 programs addressing the persons deprived of liberty. Distance learning methods are introduced both in terms of formal and informal education programs.

Garcia and Ingason (2008) mention two projects carried out in prisons in Spain, based on e-learning methods: "Mentor Classroom project" is an e-learning project which ensures great flexibility in learning, being focused on the assimilation of practical work skills, students ultimately obtaining a certificate of professional qualification, "Bip Bip Classrooms" is a program whereby persons with high risk of exclusion receive reintegration assistance through work. One of the objectives of this project is to promote digital learning.

The same purpose has the VEPS project, developed as a Grundtvig partnership between seven European countries: Greece, Denmark, Ireland, Sweden, United Kingdom, Bulgaria, and Czech Republic (2007/2008). The project seeks to strengthen the role of the educational policies in prisons, in accordance with the recommendation no. R (89) 12, Education in Penitentiaries, the Council of Europe. The project has built a website which serves as the European Virtual School Penitentiary, having as members a number of 35 European countries. The creation of such a platform that facilitates exchanges between professionals working in prisons, regarding different materials, articles, tools can also encourage the development of this innovative process in other countries. The implementation of modular systems and of a distance learning system for prisoners is also projected, which would make the learning process more flexible and individually oriented.

The ELBEP project (Eliminating Language Barriers in European Prisons through Open and Distance Education Technology) developed as a partnership between Turkey, Poland, Netherlands, Belgium, Germany, Britain and Greece has the goal of promoting the intercultural dialogue and of meeting the prison employees' communication needs with reference to the foreign prisoners (Russian, Polish, Spanish, Greek and Turkish) imprisoned in those countries. The project is focused on the five languages, as a large percentage of foreign prisoners in those countries speak these languages. Through distance learning (Open and Distance Learning), the persons working with detainees will also acquire the beginner level of their native language, thus some communication problems will be solved. The project results are reflected in the five- language A1 portals, the proposed model including e-courses, e-books, e-exercises, e-exams, e-counselling, usable starting from November 2009.

The project "Prison Knowledge-The Knowledge Community Portal for Penitentiary Institution" developed by University of Bremen offers a Prison Portal with a complex structure showed in the images below [Fri08]:

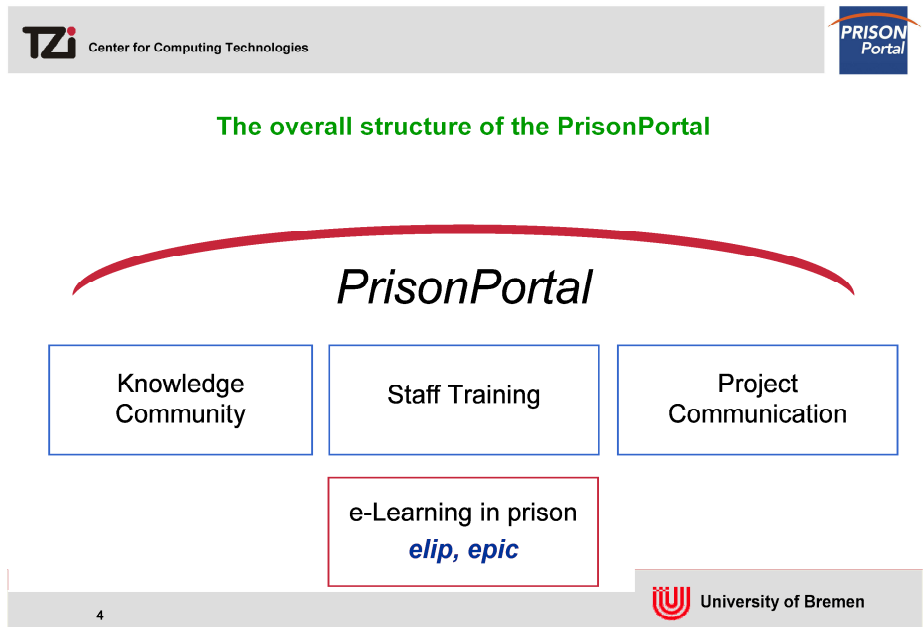

**Figure 1. The overall structure of the Prison Portal**

13

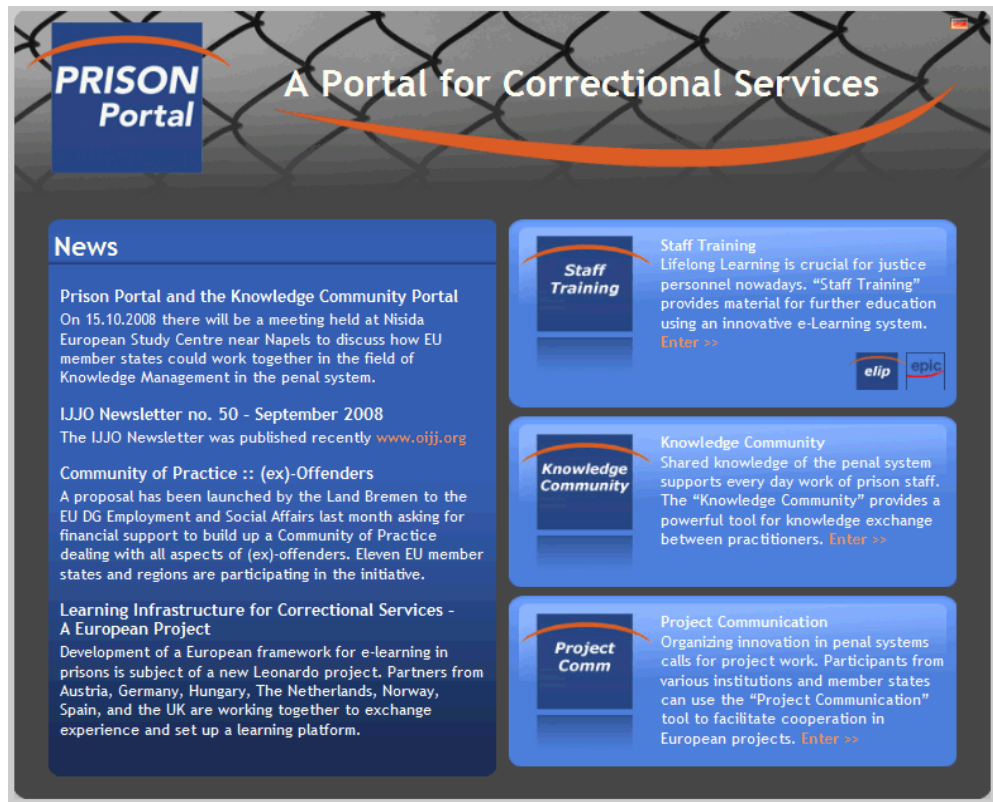

**Figure 2. Prison Portal**

In the Romanian Penitentiary System such initiatives are developed in collaboration between four institutions: Buziaş Reeducation Centre (Romania), Casa di Reclusione-ICATT (Italy), Praxiling University Montpellier3 UMR 5267 CNRS (France), Dipartimento di Sociologia e-policy -Università della Scienza degli Studi di Salerno (Italy). The general objective of the project " E-learning Education for Prisoners and Prisoners-EEPP professionals" is to establish and develop a dialogue between the penitentiary staff and the academic staff from educational organizations in terms of the work with prisoners, based on the dynamics and on the experience of using the distance education system in reducing the risk of exclusion for the prisoners and the continuous education for the penitentiary staff.

The project aims also at building a community of trainers able to address this issue and at providing new learning opportunities and reintegration in the society of persons at risk of social exclusion.

The participants in this project will gain expertise in using ICT and e-learning modalities and they will make up "The guide of the design of the on online learning in prison". This will be done by building a site for the project, through the Moodle platform. Project sustainability will be ensured by the community that will exist after the project is completed, representing the starting point for future collaboration.

**Conclusions**

The benefits of the e-learning for prisoners (Friedrich, 2008):
- broadening the areas of professional skills from the traditional ones developed in prison (mechanics, carpentry, painting) to those necessary in the most required areas (IT, industry, etc.)
- support for short term qualification (modular training);
- the possibility of individualizing the learning process (open learning environment, individual support offered by teachers, tele-tutor);
- the opportunity of developing co-operative learning (intra / inter-prison virtual classes);
- the problem-oriented learning (e.g. "discovery learning");
- high level of motivation, self-esteem, social prestige;
- integrative approach.

Due to the particularities of the work in penitentiary establishments the implementation of an e-learning system requires specific security measures (Friedrich, 2008):
- firewalls, local and central;
- Virtual Private Networks (VPN), digital signature;
- selected access to internal messaging function, logging of message content;
- moderated forum communication;
- passwords (considering biometrical identification);
- URL Filtering by using white list;
- training of personnel, cross-check.

.